# Terahertz-Driven Nanotip Field-Emission Electron Gun and Cascaded Acceleration

Wentao Yu[1,2], Nongchao Tan[1,2,3] *, Kai Peng[1,2], Sijie Fan[1,2], Yixiao Fu[1,2], Kai Jiang[1,2], Zhao Yun[1,2], Longding Wang[1,2], Renkai Li[1,2], Yingchao Du[1,2], Lixin Yan[1,2], Chuanxiang Tang[1,2],Wenhui Huang[1,2] †

[1]Department of Engineering Physics, Tsinghua University, Beijing 100084, China

[2]Key Laboratory of Particle & Radiation Imaging (Tsinghua University), Ministry of Education, Beijing 100084, China

[3]College of Science, National University of Defense Technology, Changsha 410073, China

**Corresponding authors**

*Nongchao Tan, e-mail: tannongchao@nudt.edu.cn, tel: +8615573160675, address: College of Science, National University of Defense Technology, Changsha 410073, China

† Wenhui Huang, e-mail: huangwh@mail.tsinghua.edu.cn, tel: +8613661391696, address: Department of Engineering Physics, Tsinghua University, Beijing 100084, China

**Abstract**

This paper reports two versions of terahertz (THz)-driven nanotip field-emission electron guns: single-layer reflective guns (SLRGs) and double-layer reflective guns (DLRGs). Both guns use nanotip emitters and accelerate electrons through the electric field of the THz wave. SLRGs employ a reflective structure to superimpose the initial and subsequent half-cycles of the THz electric field, enhancing the field amplitude and acceleration efficiency. Experiments have demonstrated that SLRGs achieve higher acceleration efficiency than single-layer nonreflective guns (SLNRGs) for identical THz input energies. This constitutes direct experimental verification of the efficacy of the reflective structure. Theoretically, SLRGs operating in single-feed mode can match the acceleration efficiency of dual-feed SLNRGs while reducing operational complexity. DLRGs demonstrate THz-driven cascaded electron acceleration through precise scanning of the delay between two incident THz beams. This represents a direct experimental demonstration of cascaded acceleration in THz-driven electron sources. The experimental results of DLRGs align closely with the results of electron dynamics predicted by simulations. This establishes the foundation for developing multilayer high-acceleration-efficiency THz-driven high-energy electron guns. The ability to manipulate the THz for each layer individually holds promising potential for improving the beam quality of THz electron guns.

Next-generation accelerator facilities, such as free-electron lasers (FELs) and ultrafast electron diffraction (UED), have emerged as useful tools for probing fundamental biological and chemical processes.[1-4] Compact high-gradient accelerators are necessary to downsize these facilities. Furthermore, higher acceleration gradients help mitigate space charge effects by rapidly accelerating electrons to relativistic velocities, particularly in injectors. It is suggested that the breakdown electric field strength ($E_{break}$), which limits the maximum achievable acceleration gradient, follows $E_{break} \propto f^{1/2} \tau^{-1/6}$,[5-7] indicating proportional scaling with the square root of frequency ($f$) and inverse scaling with the sixth root of pulse duration ($\tau$).

Compared with conventional microwave acceleration, terahertz (THz) radiation inherently features higher frequencies and shorter pulse durations, consequently resulting in a higher breakdown field strength threshold. This enables acceleration gradients to reach GV/m levels. Within all-optical compact desktop beamlines,[8,9] laser-driven THz sources offer intrinsic synchronization via shared seeding lasers.[10] In addition, unlike dielectric laser acceleration,[11] THz acceleration schemes leverage submillimeter structures manufactured by conventional micromachining to accelerate picocoulomb-scale bunch charge, demonstrating advantages in processing convenience and charge-carrying capacity.

Although THz-driven acceleration and beam control have been well demonstrated,[12-16] THz-driven guns for all-optical beamlines require further development. Most THz-driven electron-gun architectures implement tapered rectangular waveguide structures integrated with planar cathodes.[10,17-20] The tapered waveguide enhances free-space-to-waveguide coupling efficiency and amplifies THz electric fields. Although planar cathodes yield electron beams with low transverse emittance, their inherently lower field enhancement factors compared with those of nanotip cathodes limit the initial electron acceleration. Consequently, electrons traverse a shorter distance during acceleration and subsequently encounter a decelerating field, reducing acceleration efficiency.

Inspired by THz-driven field emission from metal tips,[21-25] a promising approach involves combining nanotips with tapered waveguide structures. In this scheme, the local electric field is enhanced to emit electrons, which then experience subcycle acceleration[26] to rapidly reach the initial velocity. Experiments on the THz-driven dual-feed photogun and the field-emission gun, which combine micrometer-scale tip cathodes with waveguide structures, have demonstrated improved acceleration efficiency through the use of this combination.[27,28] However, two THz waves are indispensable for the dual-feed operation mode, which in turn gives rise to THz power balance and synchronization issues. In addition, the energy of electron beams from THz electron guns remains relatively low. Therefore, validating cascaded acceleration on THz electron sources is necessary to realize high-energy THz electron guns.

In this work, we present two versions of reflective THz-driven nanotip field-emission electron guns: single-layer reflective guns (SLRGs) and a double-layer reflective gun (DLRG). The guns incorporate not only tapered waveguide couplers integrated with nanotip emitters but also metal shorting planes positioned at the waveguide terminations to achieve electric field enhancement via constructive interference of reflected waves. For comparison purposes, a single-layer nonreflective gun (SLNRG) was also developed, used solely in a supporting capacity. By analyzing electron energy spectra from SLRGs with varying reflective lengths, we experimentally confirm the efficacy of the structure in enhancing acceleration efficiency. Theoretically, this optimized performance can match the dual-feed operation mode when specific waveforms are employed. We developed the DLRG on the basis of the optimal SLRG design, supporting the experimental demonstration of cascaded acceleration in THz-driven electron sources.

In experiments, quasi-single-cycle THz waves were generated via optical rectification in a lithium niobate (LN) crystal and directed toward a vacuum chamber for experimentation[29] (details in the Supplementary Material). In the SLRG experimental setup (Fig. 1a), the THz wave was focused by an off-axis parabolic (OAP) and coupled into the electron gun. The full width at half maximum (FWHM) of the measured THz focus (Fig. 1c) was ~1 mm. The SLRG consisted of a rectangular waveguide with a field-emission tip centered on the wide side and a tapered rectangular horn for free-space THz coupling. The uniform rectangular waveguide section terminated with a metallic plane (short-circuit load). This termination reflected the initial half-cycle of the THz electric field, enabling constructive interference with the succeeding half-cycle to increase the field amplitude (Fig. 1e). The measured THz signal (Fig. 1d) had a main peak at ~4 ps with two side peaks of opposite polarity. We ensured that the main peak field decelerated electrons before reflection and subsequently accelerated electrons after reflection from the short-circuit load to demonstrate the effect of this reflective structure (see the Supplementary Material). The DLRG was composed of two SLRG structures with the tip emitter in the first layer (Fig. 1b). The two layers in our DLRG had different tapered waveguides with opposite coupling directions, thereby allowing each layer to be fed with different THz pulses. Consequently, the DLRG can achieve various manipulating functions by adjusting the delay of the THz wave for the second layer. In the DLRG experimental setup (Fig. 1b), a high-resistivity silicon plate served as a THz beam splitter, generating two THz pulses for the two layers. The structure was also compatible with independently generated THz pulses. During the experiment, first, the synchronization of the two THz pulses was determined (see the Supplementary Material). Then, for the cascaded acceleration of the DLRG, the delay was precisely rescanned on the basis of the initial synchronization and the electron energy spectrum. The average emission charge was measured using a Faraday cup connected to a Keysight B2985A electrometer. The electron energy spectra were acquired through retarding field analysis. Additional experimental details, including the experimental setup and gun configurations, are provided in the Supplementary Material.

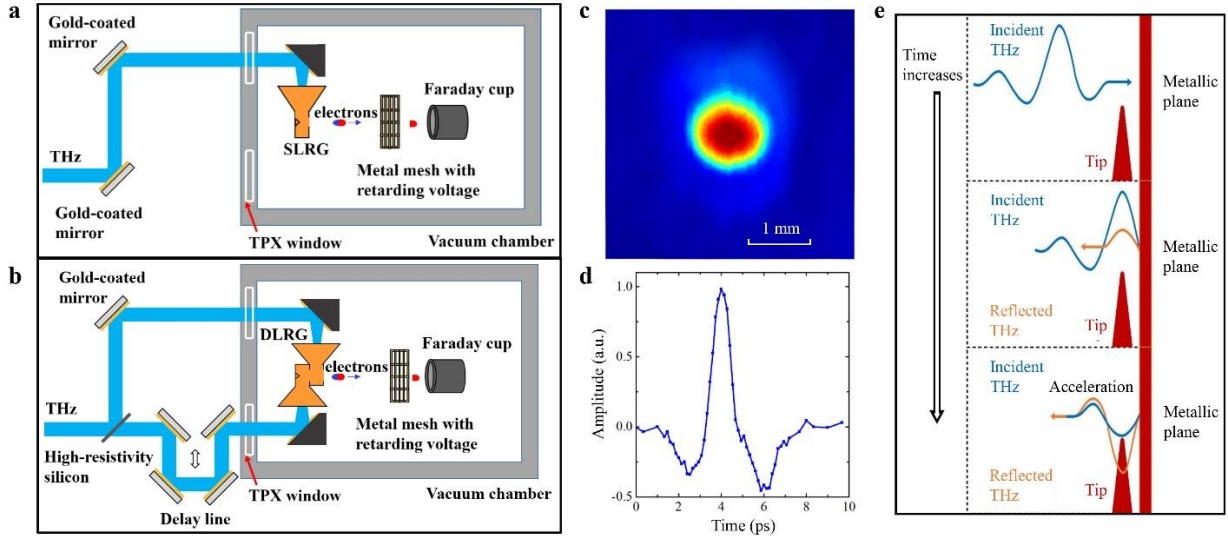

**Fig. 1.** Experimental setup (a) for the SLRG and (b) for the DLRG. (c) THz focus spot and (d) measured THz signal. (e) Schematic diagram of THz reflection superposition.

We conducted experiments on SLRGs with different reflective lengths under an input THz energy of 3 μJ to directly observe the acceleration efficiency improvement of SLRGs. The corresponding results, shown in Fig. 2 alongside the SLNRG results for comparison, demonstrated good agreement with the simulations. The field enhancement increased with increasing reflective length within the range below the optimal reflective length. Specifically, when driven by THz waveforms characterized by a sufficiently strong deceleration half-cycle, the optimized SLRG achieved an acceleration efficiency of 3.349 keV/μJ$^{0.5}$. This efficiency surpassed that of the single-feed SLNRG. Theoretically, the SLRG can achieve efficiency levels equivalent to those of a dual-feed structure without the need for dual feeds. This efficiency equality arises under specific conditions (Fig. 2c): coupling a single-cycle THz pulse whose waveform features an initial deceleration half-cycle matching the amplitude of the subsequent acceleration half-cycle into the SLRG; and simultaneously coupling two THz pulses, each with an accelerating half-cycle, into the dual-feed structure. Despite equal total coupled THz energy, the SLRG requires only a single THz wave and eliminates synchronization difficulty compared with the dual-feed SLNRG. Future work can focus on two primary areas to further enhance the performance of the structure: the generation of suitable THz waveforms and the development of an adjustable shorting plane capable of forming efficient interference superposition on the cathode across a range of THz pulse waveforms.

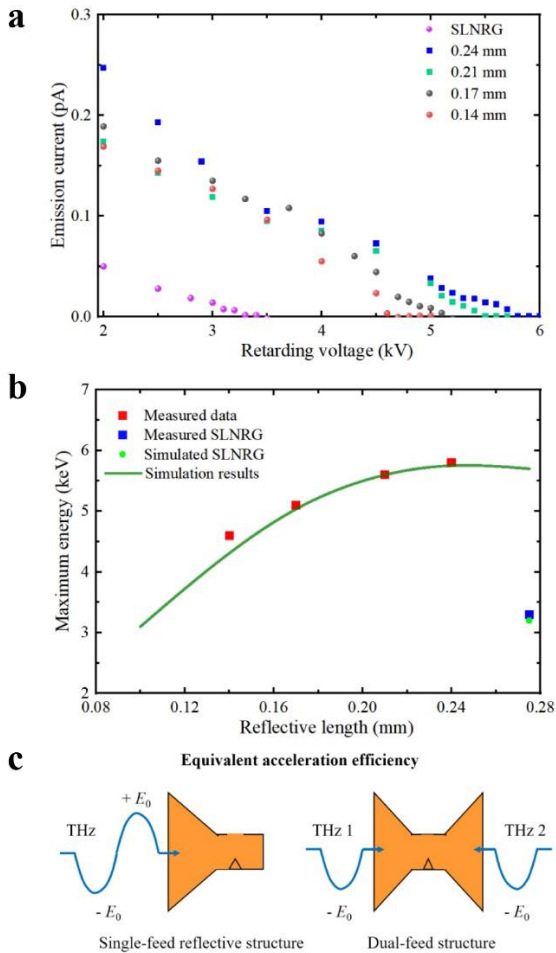

**Fig. 2.** Results of SLRGs with different reflective lengths under an input THz energy of 3 μJ. (a)

The measured energy spectra of SLRGs with reflective lengths of 0.14 mm (red spheres), 0.17 mm (gray spheres), 0.21 mm (green squares) and 0.24 mm (blue squares). The pink spheres represent the energy spectrum of the SLNRG. (b) Simulated (green curve) and experimental (red squares) maximum electron energies in SLRGs. The SLNRG results are plotted as blue squares for the experimental measurements and green circles for the simulation results. (c) Schematic diagram of the condition where the SLRG can achieve acceleration efficiency levels equivalent to those of a dual-feed structure.

Despite the nonideal THz waveform in our experiments, the peak acceleration efficiency of 3.349 keV/$\mu$J$^{0.5}$ achieved by our SLRG is comparable to the best result reported for a THz photogun with a micrometer-scale tip, operating in dual-feed mode,[27] which is 3.5 keV/$\mu$J$^{0.5}$. Although this efficiency is currently lower than that of the highest-performing THz-driven field-emission gun operating in dual-feed mode,[28] substantial scope for improvement remains with optimized THz waveforms and higher THz energies. This performance highlights the strong potential of SLRGs for developing high-gradient THz-driven electron guns.

The beam quality was evaluated by measuring the electron beam emittance via the solenoid scanning method (see the Supplementary Material). However, the maximum solenoid current was restricted by its thermal limitations in a vacuum, which in turn constrained the maximum electron energy to ~2.8 keV to ensure proper beam focusing on the screen. By systematically varying the applied solenoid current, the beam spot sizes were measured on the screen, as shown in Fig. 3.

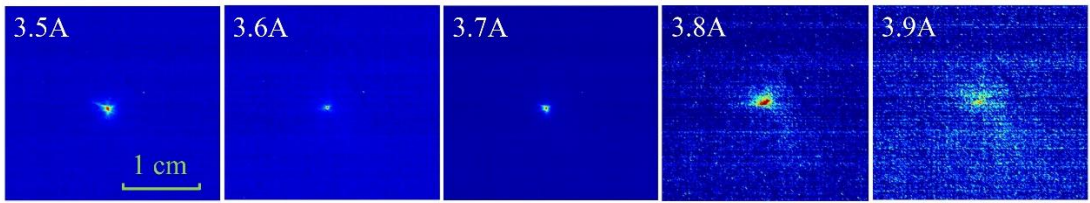

**Fig. 3.** Beam spot sizes at different solenoid currents.

The fitted normalized emittance from the experimental data was 1.42 $\pi$ mm·mrad. The emittance values were nearly identical in the two transverse directions, which is not typical for a slit beam exit. We attributed this symmetry to the large angular divergence of the beam emitted from the tip, causing the beam to be constrained by the solenoid. Our simulation successfully captured this behavior, predicting a final beam normalized emittance of 1.66 $\pi$ mm·mrad and 1.36 $\pi$ mm·mrad in these two transverse directions. Despite the small emission area, the nonlinear processes resulting from the large angular divergence led to a significant increase in the emittance during beam transport.

In addition to the SLNG structure, we developed a DLRG structure as well. Although extensive simulation studies have been conducted for multilayer THz electron guns,[18,27,28,30] direct experimental validation of THz cascaded acceleration in THz-driven electron sources is still lacking. Moreover, experiments on multilayer cascaded acceleration typically employ integrated structures with timing-controlled dielectric sheets, precluding direct characterization of individual layer contributions. For an assembled structure, adjusting the THz phase and energy of different layers individually is not possible. In our DLRG experiment, we directly verified the role of individual layers and achieved an experimental demonstration of cascaded acceleration in a THz-driven electron source. After precisely scanning the delay between the two THz beams and simultaneously measuring the peak electron energy, we obtained the optimal cascaded acceleration and deceleration results, as shown in Fig. 4(a). The electron energy spectra are shown for three conditions in the second layer: no THz excitation (black spheres), THz-driven cascaded acceleration (red spheres), and THz-induced deceleration (green spheres).

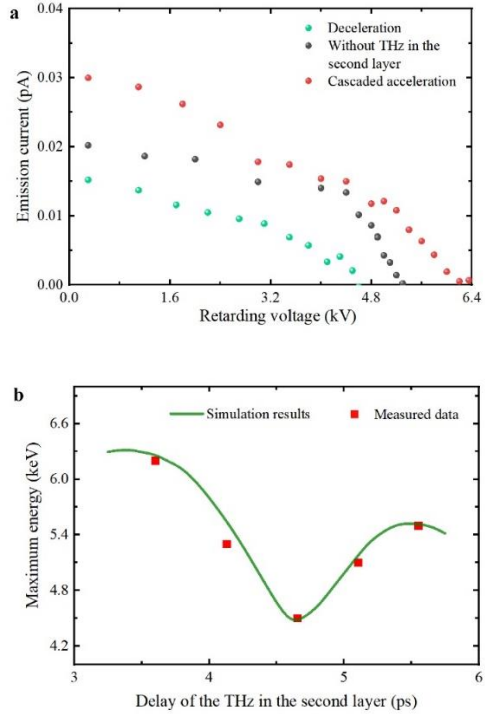

**Fig. 4.** The electron energy spectrum results of the DLRG. (a) The optimal cascaded acceleration (red spheres) and deceleration (green spheres) results of the DLRG. The black spheres represent the results without the THz beam in the second layer. (b) The maximum energy of the electron beam as a function of the delay of the two THz beams. The green curve represents the simulation results, and the red squares represent the measured data.

Compared with the initial synchronization (where the two THz pulses arrive at the cathode almost simultaneously; see the Supplementary Material), in the optimal cascaded acceleration mode, the delay of the THz beam in the second layer relative to the THz beam in the first layer was ~3.6 ps. This corresponds to the time it takes for the 5.2 keV electrons in the first layer to traverse the slit on the intermediate metal sheet with a thickness of 100 μm and then encounter the onset of the acceleration field in the second layer.

Owing to the actual split THz power ratio (first layer: 85%, second layer: 15%) (see the Supplementary material), the maximum electron energy increased from 5.2 keV to 6.2 keV in the cascaded acceleration mode and decreased from 5.2 keV to 4.6 keV in the deceleration mode. Since the input THz energy to the second layer was suboptimal, electrons traversing the second layer experienced both acceleration and deceleration, falling short of the initial design expectations. Nevertheless, the simulated dynamics using the actual split power ratio remained largely consistent with the experimental observations, as shown in Fig. 4(b), with ~3 μJ input for the first layer and ~0.53 μJ input for the second layer. Notably, the constraint of the split THz power ratio was determined by our method to generate two THz pulses (using a high-resistivity silicon plate as the THz beam splitter), which is unrelated to the structure of the electron gun. The midpoint between the acceleration peak and the deceleration valley is believed to be a zero-crossing phase, where the electron energy spectrum can be compressed or broadened. However, these phenomena were obscured because of the large energy spread of the electrons. Although this limitation prevented the observation of finer details, the DLRG experiment achieved its primary objectives. This experiment successfully provided direct verification of the individual role of each layer and experimentally demonstrated THz-driven cascaded acceleration from THz electron sources.

In conclusion, this study focused on THz-driven electron guns that combine tapered waveguide structures with field emission from nanotips. We conducted experimental investigations on two versions of these guns: an SLRG and a DLRG. SLRGs employ a reflective structure to generate constructive interference between the reflected initial half-cycle of the THz electric field and the subsequent half-cycle, thereby increasing the overall field amplitude. Compared with SLNRGs, SLRGs experimentally exhibit higher acceleration efficiency under identical THz input energies. This single-feed structure, which uses a single-cycle THz waveform with balanced positive and negative amplitudes, can theoretically achieve an acceleration efficiency comparable to that of a dual-feed, nonreflective structure driven by half-cycle THz waveforms. Furthermore, it eliminates the requirement for precise time synchronization between two THz beams, highlighting its promising competitiveness for realizing high-gradient THz-driven electron guns. The DLRG experiment demonstrated THz-driven cascaded acceleration from THz-driven electron sources. Its unique double-layer structure enables various manipulating functions via changing the delay of the THz waves. The electron dynamics were also accurately replicated by our simulations. These results establish the groundwork for developing high-energy, multilayer THz electron guns. Future work will focus on enhancing the THz waveform quality, increasing the THz pulse energy, and optimizing the accelerating structure to generate electron beams of higher energy and superior quality. We believe that this work paves the way for the development of all-optical, desktop-compact beamlines for applications in ultrafast electron diffraction, ultrafast electron microscopy, and attosecond X-ray sources.

## Acknowledgments

The authors acknowledge support from the National Natural Science Foundation of China (NSFC Grant No. 12035010 and No.12405179).

## Author Declarations

### *Conflict of interest*

The authors have no conflicts to disclose.

### *Author contributions*

Wentao Yu: Conceptualization (equal); Investigation (equal); Formal analysis (equal); Methodology (equal); Writing – original draft (lead); Writing – review & editing (equal). Nongchao tan: Funding acquisition (equal); Writing – review & editing (equal); Formal analysis (equal). Kai Peng: Investigation (equal); Formal analysis (equal); Methodology (equal). Sijie Fan: Investigation (equal). Yixiao Fu: Investigation (equal). Kai jiang: Investigation (equal). Zhao Yun: Investigation (equal). Longding Wang: Investigation (equal). Renkai Li: Resources (equal); Yingchao Du: Resources (equal); Lixin Yan: Resources (equal); Chuanxiang Tang: Resources (equal); Wenhui Huang: Conceptualization (equal); Formal analysis (equal); Methodology (equal); Resources (equal); Funding acquisition (equal); Supervision (lead); Project administration (lead); Writing – review & editing (equal).

## Data availability

The data that support the plots within this paper and other findings of this study are available from the corresponding author upon reasonable request.

**Supplementary Material for**

**Terahertz-Driven Nanotip Field-Emission Electron Gun and Cascaded Acceleration**

Wentao Yu[1,2], Nongchao Tan[1,2,3] *, Kai Peng[1,2], Sijie Fan[1,2], Yixiao Fu[1,2], Kai Jiang[1,2], Zhao Yun[1,2], Longding Wang[1,2], Renkai Li[1,2], Yingchao Du[1,2], Lixin Yan[1,2], Chuanxiang Tang[1,2], Wenhui Huang[1,2] †

[1]Department of Engineering Physics, Tsinghua University, Beijing 100084, China

[2]Key Laboratory of Particle & Radiation Imaging (Tsinghua University), Ministry of Education, Beijing 100084, China

[3]College of Science, National University of Defense Technology, Changsha 410073, China

**Corresponding authors**

* Nongchao Tan, e-mail: tannongchao@nudt.edu.cn, tel: +8615573160675, address: College of Science, National University of Defense Technology, Changsha 410073, China

† Wenhui Huang, e-mail: huangwh@mail.tsinghua.edu.cn, tel: +8613661391696, address: Department of Engineering Physics, Tsinghua University, Beijing 100084, China

*THz generation and measurement*

The setup for THz generation, transmission, and characterization is shown in Fig. S1. A Ti:sapphire laser system was used as the pump laser with a maximum output energy of 20 mJ, a repetition rate of 10 Hz, and a pulse duration of ~800 ps at a central wavelength of 800 nm. The pulse is compressed by a dual-grating system to approximately 400 fs and passes through a beam splitter. The reflected pulse serves as a probe for THz electric field electro-optic measurements. The transmitted pulse is diffracted by a grating to create a tilted wavefront, which is then used to generate the THz wave via optical rectification in the lithium niobate (LN) crystal.

The THz radiation emitted from the LN crystal is processed by two gold-coated mirrors to facilitate the subsequent electron gun beamline configuration. This arrangement adjusts the spatial orientation of the THz radiation while simultaneously converting its polarization from vertical to horizontal. For efficient long-distance transmission, the THz radiation is collimated into a parallel beam by a pair of gold-coated OAP mirrors. THz characterization is performed using a holed OAP mirror on an electronic displacement platform. The energy is measured with a Golay cell; the focal spot profile is acquired with an INO THz camera; and the time-domain signals are detected via a ZnTe crystal and balanced photodiodes. Additional details can be found in Ref 29.

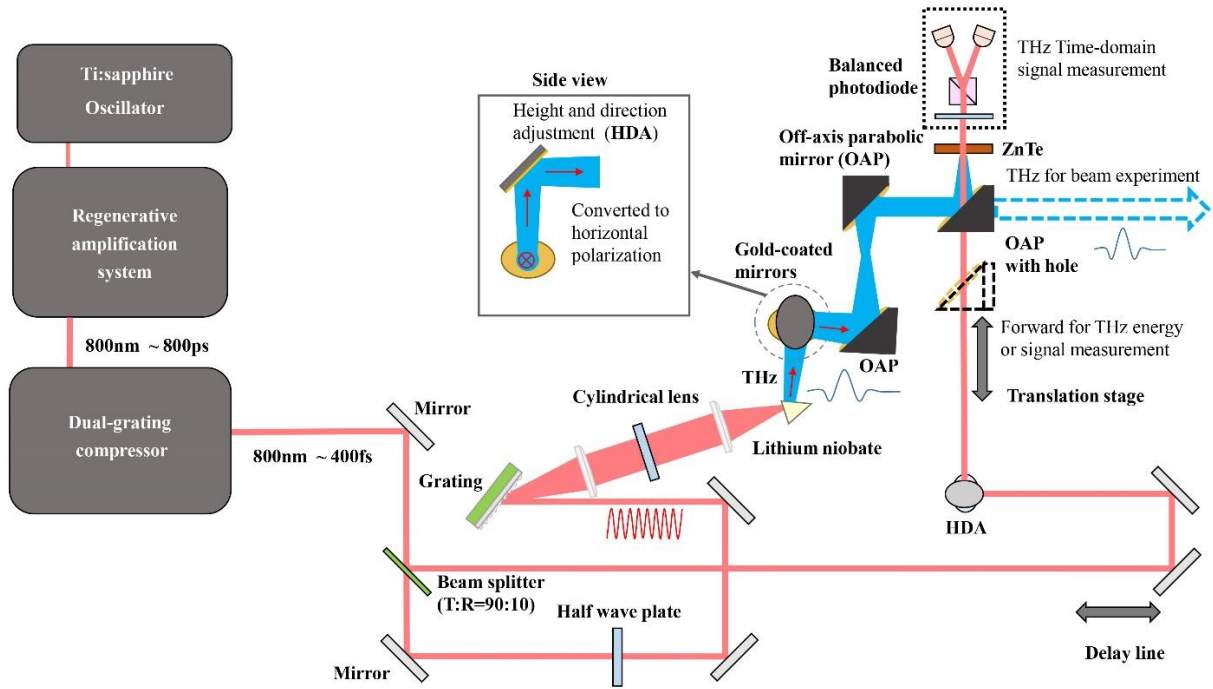

**Fig. S1.** Schematic of THz pulse generation and measurement.

### *Structural design and optimization*

This section details the structural design and simulation optimization for electron guns driven by quasi-single-cycle THz waves with a 300 GHz central frequency. The guns operate in $TE_{10}$ mode. The transverse electric field within the structure induces electron emission and subsequently accelerates the emitted electrons. Oxygen-free copper serves as the fabrication material.

### *Single-layer nonreflective gun (SLNRG)*

The SLNRG structure consists of two tapered couplers and a uniform rectangular waveguide for interaction. The two symmetrically designed tapered couplers enhance THz radiation coupling from free space into the small uniform rectangular waveguide. This configuration was originally conceived for developing a dual-feed electron gun. The dual-feed design is used to determine the main peak electric field direction within the structure and to establish synchronization between the split THz pulses. A detailed description of these processes is provided later. The schematic parametric model and the photograph of the fabricated SLNRG are shown in Fig. S2.

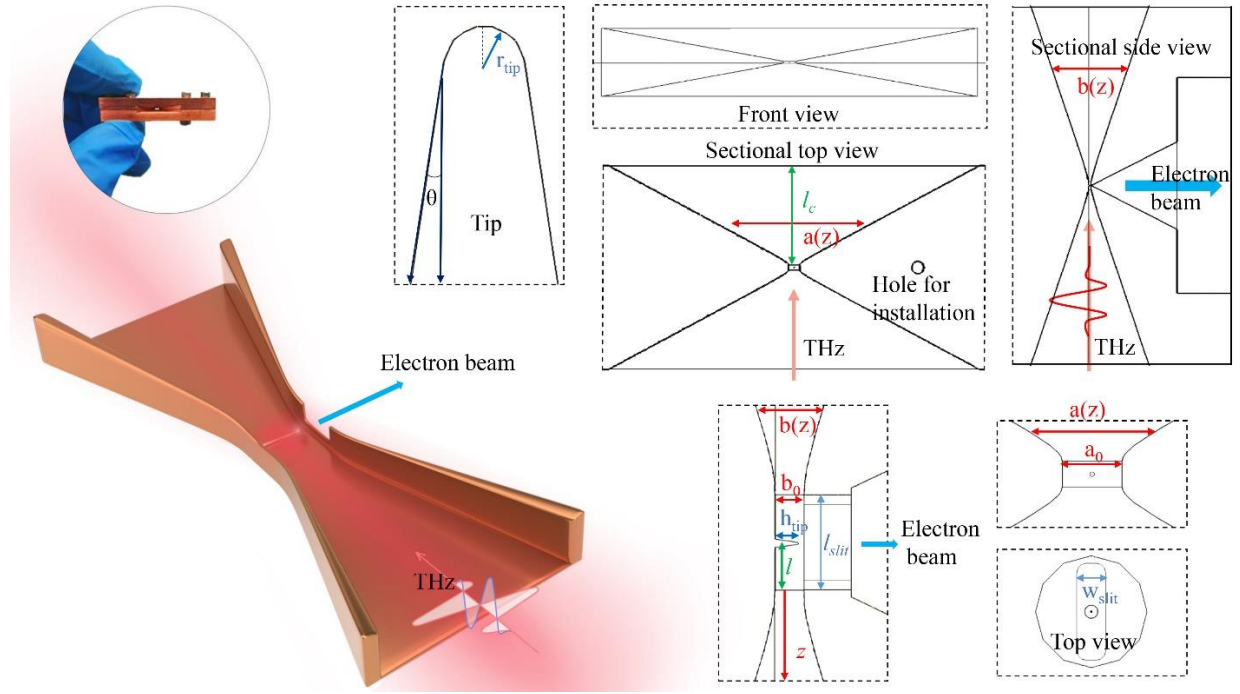

**Fig. S2.** Schematic parametric model and photographs of the SLNRG.

The two couplers are tapered rectangular waveguides. The width of the wide side is denoted as $a(z)$, and the narrow side is denoted as $b(z)$, where $z$ represents the distance from the uniform waveguide region in the coupler. The variation in $b(z)$ follows a functional form similar to a Gaussian beam profile:

$$b(z) = b_0 \sqrt{1 + (\frac{z}{z_{rb}})^2} \tag{1}$$

Here, $b_0$ represents the narrow-side length of the uniform waveguide, and $z_{rb}$ is a characteristic parameter governing the taper profile. To maintain a constant characteristic impedance along the tapered waveguide for suppressing THz reflection during coupling, $a(z)$ satisfies the following equation:

$$\frac{b(z)}{\sqrt{a(z)^2 - (\frac{\lambda}{2})^2}} = \frac{b_0}{\sqrt{a_0^2 - (\frac{\lambda}{2})^2}}$$ (2)

Here, $a_0$ is the wide-side width of the uniform waveguide, and $\lambda$ is the central wavelength of the THz radiation (approximately 1 mm in our experiments). The coupler length $l_c$ should be sufficiently long for a given taper parameter $z_{rb}$ to satisfy two key conditions: (1) $a(l_c) \gg \lambda$; (2) $b(l_c)$ must exceed the THz beam spot size at the coupler entrance. The requirement of $a(l_c) \gg \lambda$ ensures that the wave impedance at the coupler entrance approaches the vacuum wave impedance, thereby minimizing reflection when coupling THz radiation from free space into the waveguide. The requirement for $b(l_c)$ ensures that the coupler entrance accommodates nearly the entire THz beam spot size, enhancing the energy collection efficiency. The optimized parameters yield $l_c$=4.00 mm and $z_{rb}$=0.0886 mm.

The uniform waveguide region is mirror symmetrical, with its structural parameters optimized specifically for electron acceleration efficiency. These parameters are as follows:

➢ **Inner transmission distance ($l$):** This parameter denotes the distance from the central tip to the entrance of the uniform waveguide, which is equivalent to half the total length of the uniform waveguide. With consideration of tip fabrication, $l$ is subsequently optimized to be 0.1 mm. This design is beneficial for suppressing dispersion and enhancing peak field strength, simultaneously facilitating the fabrication of the tip.

➢ **Waveguide wide-side length ($a_0$):** This parameter defines the wide-side length of the rectangular waveguide. Reducing the wide-side length confines the THz energy more tightly, increasing the field strength. However, excessive reduction causes severe cutoff of low-frequency components in the broadband THz spectrum. Hence, $a_0$ is ultimately optimized as 0.47 mm, where the central frequency is below the cutoff frequency. This occurs because the transmission distance $l$ is sufficiently short such that some frequency components just below the cutoff can achieve adequate field penetration, allowing them to interact effectively with the tip.

➢ **Waveguide narrow-side length ($b_0$):** This parameter represents the narrow-side length of the rectangular waveguide. Similar to $a_0$, while reducing the narrow-side length confines the THz energy more tightly to increase the field strength, excessive reduction diminishes the central field amplification because of energy dissipation through the beam aperture. This parameter is optimized to maximize the central field strength while accounting for the beam aperture size, yielding an optimal value of $b_0$ = 60 μm.

➢ **Slit aperture beam exit length ($l_{slit}$) and width ($w_{slit}$):** These parameters define the slit aperture dimensions for beam extraction from the waveguide. The slit geometry ensures a sufficiently large beam exit acceptance to allow for the investigation of beam properties, simultaneously minimizing THz energy leakage through alignment of its long side with the surface current direction in $TE_{10}$ mode. $l_{slit}$ is designed to be 200 μm, and $w_{slit}$ is designed to be 60 μm under the given fabrication constraints.

➢ **Cone emitter height ($h_{tip}$), cone angle ($\theta$), and tip radius ($r_{tip}$):** These parameters define the geometry of the emitter in the center of the structure. As shown in the representative scanning electron microscopy (SEM) image (Fig. S3), the emitter features a conical profile. This shape is fabricated by annular plasma-focused ion beam (PFIB) from a premachined cylindrical precursor with a diameter of ~70 μm. During this process, the cone angle and tip radius are determined to be ~8.5° and ~120 nm, respectively. The cone height is optimized to maximize the electron beam energy gain. Although taller cones result in higher field enhancement factors, the optimal value requires balancing this against the baseline field distribution (emitter-free) and electron transit distance within the structure. For a THz input energy of 4 μJ, the optimized height $h_{tip}$ is ~50 μm.

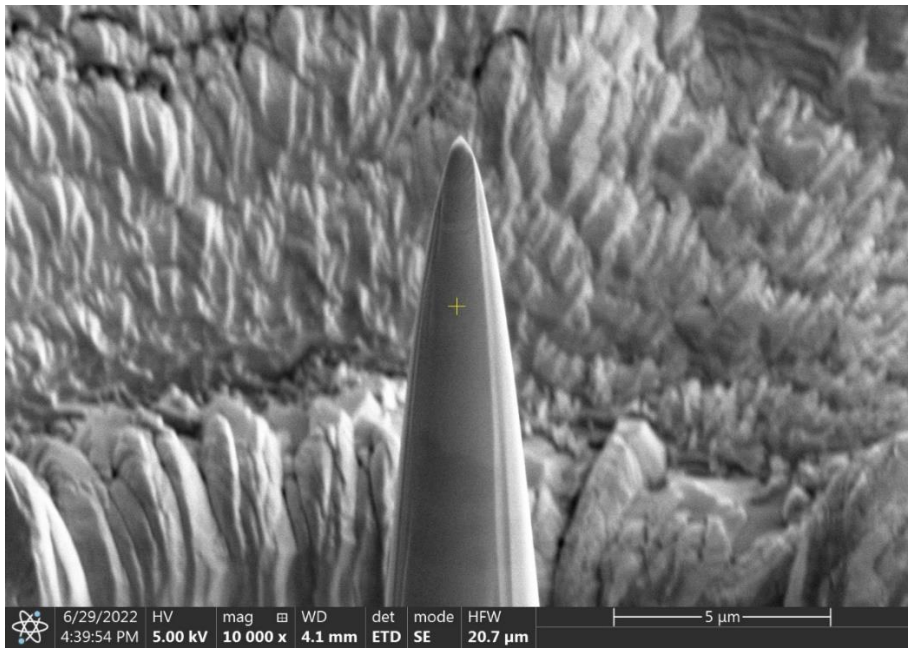

**Fig. S3.** SEM image of a representative actual tip emitter.

With these parameters, the gun model can be constructed. Future work will further optimize additional parameters, particularly the tip geometry, on the basis of the outcomes of this round of experiments.

*Single-layer reflective gun (SLRG)*

The structure of the SLRG consists of a tapered coupler and a uniform rectangular waveguide for interaction, as shown in Fig. S4. The unique parameter $l_r$ defines the distance from the emitter to the terminal reflective metal surface. A summary of the optimized SLRG parameters can be found in Extended Data Table I of Ref 29. All the structural parameters of the SLRG are identical to those of the SLNRG, except for $l_r$. This parameter determines the phase of the superimposed wave. Choosing an appropriate $l_r$ allows the reflected former half-cycle to constructively superpose with the subsequent half-cycle, enhancing the field at the tip. We optimized this reflective distance via simulations, determining an optimal $l_r$ value of 0.24 mm for our structure. To experimentally verify this, we fabricated SLRGs with different $l_r$ values from 0.14 mm to 0.24 mm, alongside a nonreflective structure (a traversed structure, which is equivalent to approximating $l_r$ as infinitely long). This nonreflective structure serves as a comparison baseline for the results of SLRGs, demonstrating the superior field enhancement capability and higher acceleration efficiency of the SLRGs.

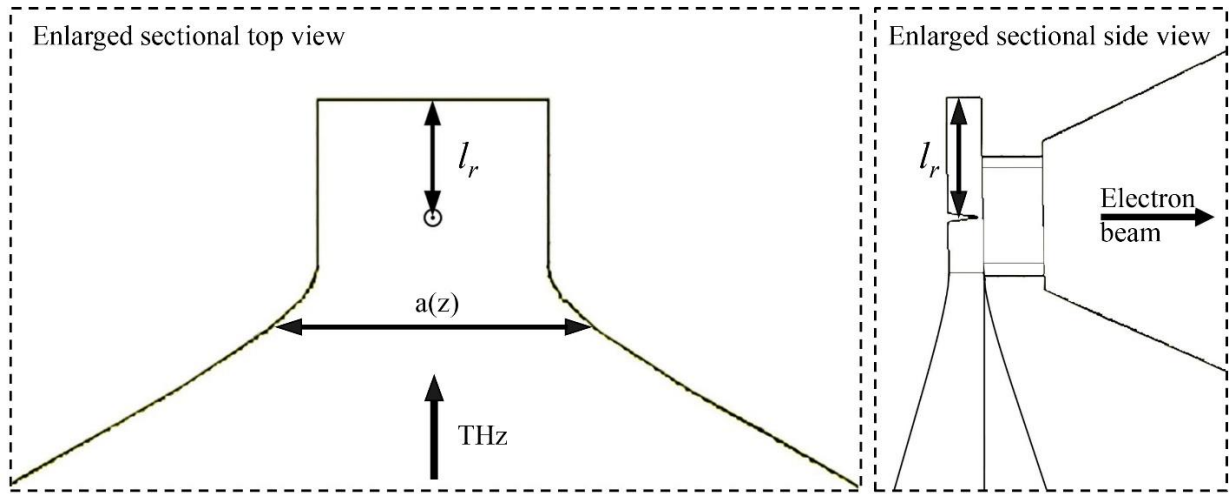

**Fig. S4.** Schematic model of the SLRG.

*Double-layer reflective gun (DLRG)*

The DLRG is composed of two layers of structures similar to the SLRG, a schematic model of which is shown in Fig. S5.

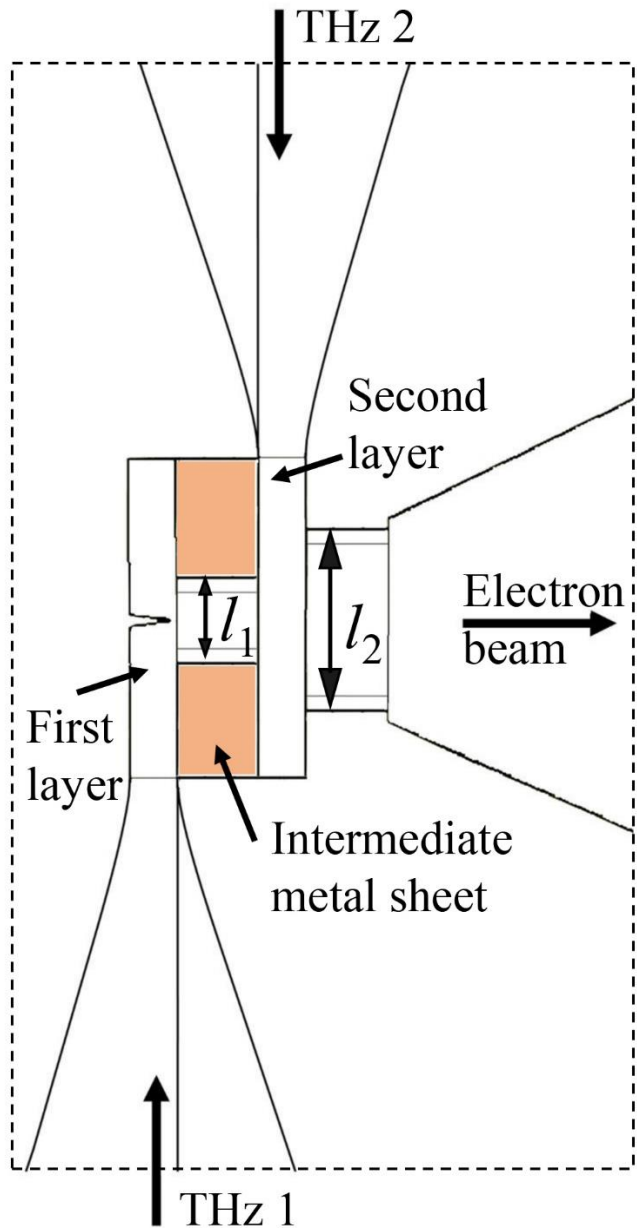

**Fig. S5.** Schematic model of the DLRG.

The two layers of the couplers are oriented in opposite directions, meaning that the THz waves coupled into each layer propagate in opposite directions. The THz waves for each layer can be generated and coupled independently, enabling direct experimental verification of the effect of the second layer. This design allows flexible tuning of the amplitude and phase of the incident THz pulse for each layer according to the dynamic design and experimental results. Therefore, a single DLRG can be configured to fulfill a variety of functions.

The designed reflective distance of the DLRG is identical to that used in the SLRG. To facilitate fabrication, the inner transmission distance $l$ is set equal to the reflective distance $l_r$. Compared with SLRGs, the extended beam path in the DLRG leads to greater electron interception by the structure when an aperture of the same size is used. We optimized three key parameters to generate a measurable output electron beam while suppressing significant field leakage: the thickness of the intermediate metal sheet, the slit aperture on the intermediate metal sheet, and the exit slit aperture of the second layer. The optimized thickness of the intermediate metal sheet was 100 μm. The length ($l_1$) and width of the slit aperture on this sheet were 110 μm and 60 μm, respectively. The length ($l_2$) and width of the slit aperture of the second layer were 270 μm and 80 μm, respectively. All the other parameters remained identical to those used in the SLNRG structure.

### THz splitting ratio

In the DLRG experimental setup, a high-resistivity silicon plate served as the THz beam splitter. The relative permittivity of this material is 11.7, and the relative permeability is 1. We theoretically calculated the electric field reflection and transmission coefficient for a parallel-polarized electromagnetic wave with an incident angle of 45°. The results are shown in Fig. S6 and are consistent with the experimental measurements. Owing to multiple internal reflections and transmissions of the THz beam at the air–silicon interfaces, the silicon plate generated a sequence of multiple output THz beams. Among these, the first three generated beams carried the majority of the energy. Their respective electric field coefficients were -0.424, 0.820, and 0.345, and the other beam energies could be neglected. The negative field coefficient indicated a 180° phase shift, typically caused by reflection from an optically rarer medium to a denser medium. The transmitted beam (THz t1) was coupled into the first layer of the DLRG to generate and accelerate electrons. The reflected beams (THz r1 or THz r2) were coupled into the second layer, and one was chosen to provide cascaded acceleration by adjusting the delay of THz t1.

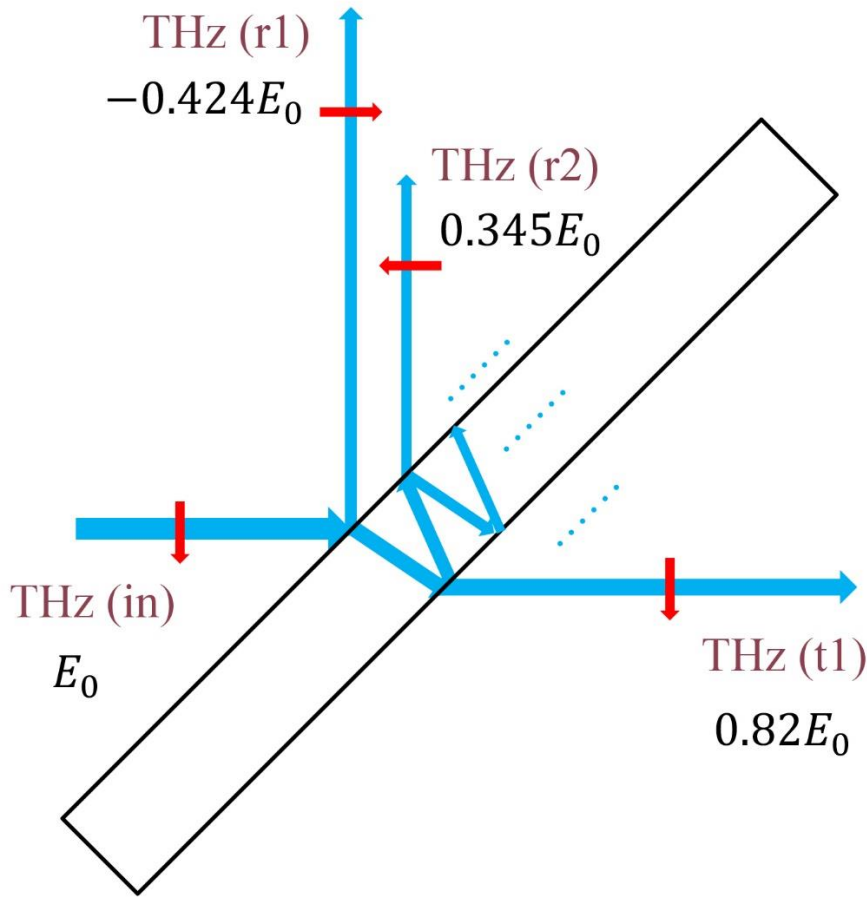

**Fig. S6.** THz wave split by a high-resistivity silicon plate.

*SLNRG experiment*

The experimental configuration for the SLNRG is shown in Fig. S7. Two CF63 TPX windows were mounted on the vacuum chamber due to their high THz transmittance. The OAP mirrors inside the chamber focused the THz waves, which were then coupled into the electron gun to trigger field emission and accelerate electrons. A Faraday cup was located downstream to measure the average emission current by a Keysight B2985A electrometer. A metal mesh biased with a negative high voltage was used to analyze the electron energy spectrum. The normalized energy spectrum could be derived by differentiating the current data with respect to the retarding voltage.

*Determination of the main peak field direction*

In the SLNRG experiment, we first adopted the dual-feed design of the SLNRG to input THz waves in two distinct ways, as shown in Fig. S7. Under identical THz energy levels, the input Path T generated electron beams with higher energy than the input Path R did, indicating that the strongest (main peak) THz electric field (Fig. 1(d), appearing at ~4 ps) pointed toward the cathode tip and accelerates electrons when Path T was applied. In the case of Path R, the strongest THz electric field pointed away from the cathode tip, and the side peaks accelerated electrons, leading to lower electron energies. However, Path R ensured that the strongest reflected THz field accelerated electrons and superimposed onto the smaller side peak. Therefore, the efficacy of the reflective structures was more pronounced with Path R than with Path T. We used Path R for all the SLRGs, including the comparative SLNRG structure.

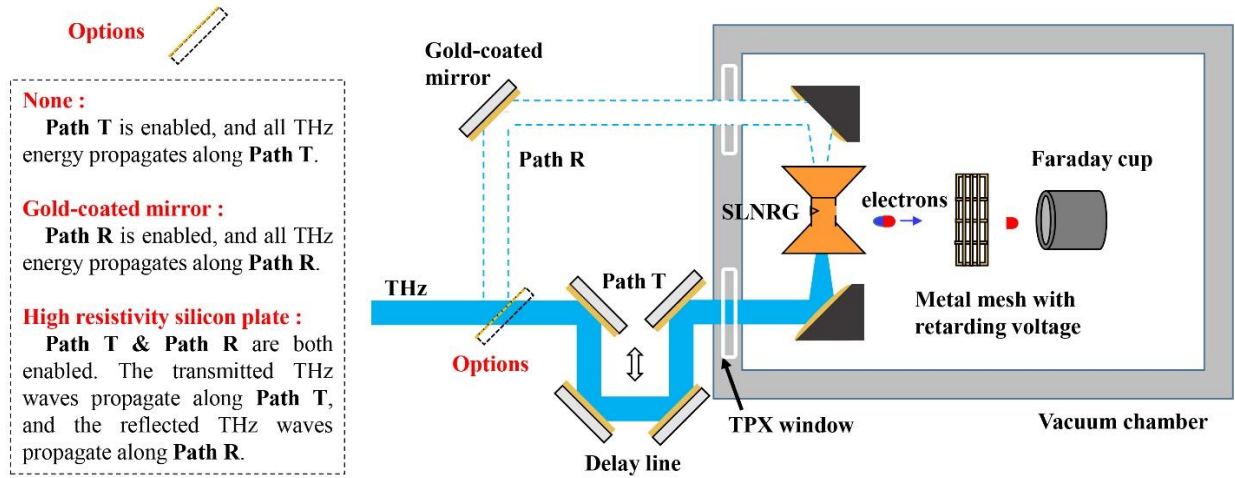

**Fig. S7.** The experimental configuration for the SLNRG.

*Determination of the initial time synchronization*

A high-resistivity silicon plate was positioned at the position of Options in Fig. S7 to synchronize the two THz beams. The THz delay line on Path T for THz t1 (Fig. S6) was activated. The SLNRG was positioned such that THz pulses from both Path T and Path R could be efficiently coupled into it. This facilitated easy identification of the initial time synchronization by monitoring the emission current of the SLNRG, where the two THz pulses arrived at the tip almost simultaneously. The emission current signals of the transmitted THz t1, interfering with the reflected THz r1 and with the later reflected THz r2, are shown in Fig. S8. They varied in the same pattern but in opposite directions, confirming our theoretical analysis. THz r2 was chosen for the cascaded acceleration experiment because it was in the same phase as THz t1, allowing the strongest main peak field to accelerate electrons. Afterward, the SLNRG was replaced by the DLRG to demonstrate the cascaded acceleration. Owing to the electron flight time and temporal continuity of the acceleration fields in the two layers, the delay was precisely rescanned to perform cascaded acceleration or other manipulations of the electron beam.

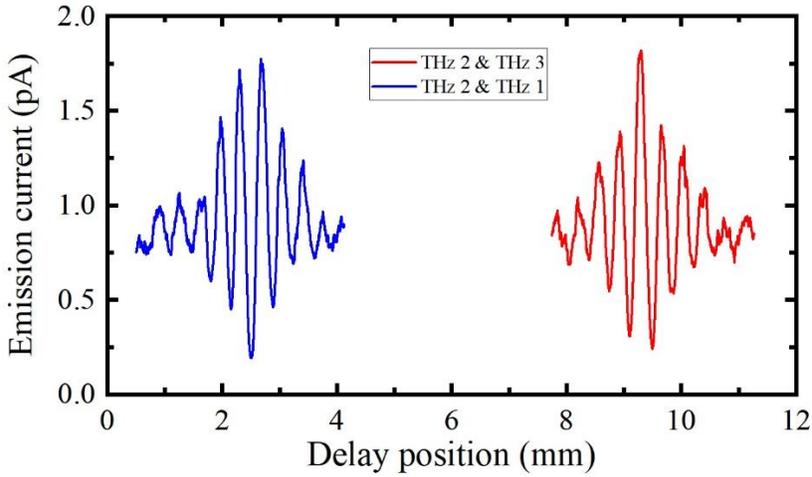

**Fig. S8.** Variation in the emission current from the SLNRG in dual-feed mode with respect to the delay position.

*Subcycle acceleration*

The experimental results of the SLNRG with the THz input Path T are presented in Fig. S9. The maximum energy of the electrons emitted at different THz pulse energies can be seen in Fig. S9 (a). Consequently, the relationship between the maximum energy of the emitted electrons and the peak field strength was derived using electron dynamics simulations, as shown in Fig. S9(b). The nearly linear dependence of the maximum electron energy on the peak field strength indicated subcycle acceleration dynamics in the SLNRG and was consistent with the simulation results. With respect to the SLRGs and the first layer of the DLRG, which have higher acceleration efficiency than the SLNRG does, subcycle acceleration dynamics are also natural consequences.

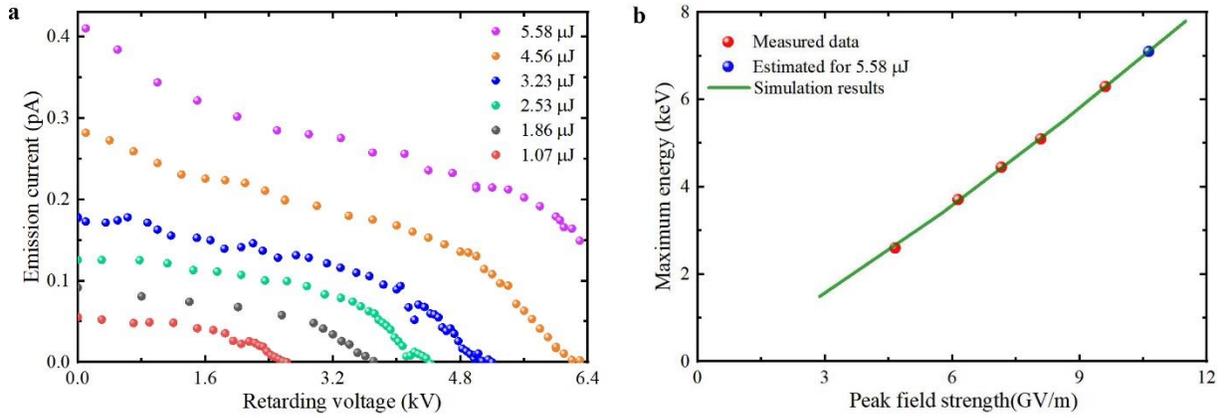

**Fig. S9.** The experimental results of the SLNRG with the THz input Way 1. (a) Energy spectra data measured by the retarding field method under different THz energies. (b) Relationships between the maximum energy of the emitted electrons and the peak field strength. The peak field strength is determined by our simulations. The THz energy is proportional to the square of the peak field strength.